\documentclass[notitlepage,12pt,onecolumn,openbib]{article}%
\usepackage{amssymb}
\usepackage{amsmath}
\usepackage{geometry}
\usepackage{amsfonts}
\usepackage{graphicx}%
\begin{document}

\title{Currents, Torques, and Polarization Factors in Magnetic Tunnel Junctions}
\author{J. C. Slonczewski$^{1}$\\IBM Watson Research Center, Box 218, Yorktown Heights,\\NY 10598 USA\\ \\Received by Phys. Rev. B on April 7, 2004. \ Revised on}
\maketitle

\begin{abstract}
Bardeen's transfer-hamiltonian method is applied to magnetic tunnel junctions
having a general degree of atomic disorder. \ The results reveal a close
relationship between magneto-conduction and voltage-driven pseudo-torque, and
also provide a means of predicting the thickness dependence of
tunnel-polarization factors. \ Among the results: 1) The torque generally
varies with moment direction as $\sin\theta$\ at constant applied voltage.
\ 2) Whenever polarization factors are well defined, the voltage-driven torque
on each moment is uniquely proportional to the polarization factor of the
other magnet. \ 3) At finite applied voltage, this relation implies
significant voltage-asymmetry in the torque. \ For one sign of voltage the
torque remains substantial even if the magnetoconductance is greatly
diminished. \ 4) \ A broadly defined junction model, called \emph{ideal
middle}, allows for atomic disorder within the magnets and F/I interface
regions. \ In this model, the spin- ($\sigma$)\ dependence of a basis-state
weighting factor proportional to the sum over general state index $p$\ of
$(\smallint\smallint dydz\Psi_{p,\sigma})^{2}$\ evaluated within the (e.g.
vacuum) barrier generalizes the local state density in previous theories of
the tunnel-polarization factor. \ 5) For small applied voltage,
tunnel-polarization factors remain legitimate up to first order in the inverse
thickness of the ideal middle. \ An algebraic formula describes the
first-order corrections to polarization factors in terms of newly defined
lateral auto-correllation scales.


\end{abstract}
\date{}

\footnotetext[1]{IBM RSM Emeritus.
\par
\ john.slonczewski@verizon.net}\pagebreak

\section{Introduction}

When first predicted, voltage-driven pseudo-torque in magnetic tunnel
junctions (MTJs) appeared to be a marginal effect \cite{'89}. \ (Sec. 2
explains our use of the prefix pseudo- in the term pseudo-torque.) \ The
lithographic scales and resistances available in early experimental MTJs
appeared too large to permit anything more than a very small torque term in
the Landau-Lifshitz equation.\ \ Resistive heating of the MTJ would have
limited its possible consequences to only a small voltage-driven decrease of
linewidth of narrowly-focussed Brillouin scattering. (This prediction was
never tested.) \ As a result, one could not yet predict anything as remarkable
as the now well-established magnetic reversal and high-frequency precession
observed when the resistive barrier is replaced by a \textit{metallic} spacer.
\ For recent experimental work and earlier references dealing with switching
and current-driven oscillations involving metallic spacers, see Refs.
\cite{Nature AC} and \cite{RipOsc}.

But in recent years, experimental activity in tunneling magnetoresistance has
expanded vastly. \ It is fueled in great part by the experimental discovery of
substantial tunneling magnetoresistance \cite{Mood} at room-temperature and
the resulting intensive exploration of non-volatile magnetic memory reviewed
recently \cite{mram}. \ A part of this activity is the search for junction
compositions and deposition techniques which lower the resistance to values
more suitable for integrated-circuit application. \ Indeed, there now exist
very recent experimental reports of current-driven switching in MTJs
\ \cite{Nguyen,Fuchs}. \ This development may make possible two-terminal
memory elements avoiding resort to three-terminal devices using both a
metallic spacer for switching and a tunnel barrier for reading \cite{FIM99}.

According to recent reviews of tunneling magneto-resistance \cite{MoodMatRev,
MiyazakiReview, MTIRev, TsymbRev}, empirical ferromagnet polarization
coefficients $P_{i}$ [$i=$L,R refer to left and right magnets F$_{i}$ in Fig.
1(a).] measured with F$_{i}$IS junctions having a superconducting counter
electrode \cite{TedMes} account well for magneto-resistance in FIF junctions.
\ Let the formula%
\begin{equation}
J(V,\theta)=-J_{0}(V)[1+\mathit{\iota}\mathcal{(}V)\cos\theta],\text{
\ with\ \ }J_{0}>0\text{ \ for \ }V>0 \label{magcur}%
\end{equation}
for current density at constant applied voltage\textit{\ }$V$\textit{\ }define
the dimensionless coefficient $\mathit{\iota}$ of magneto-conduction. \ Here
$\theta$ is the angle between the moments. \ (The $-$ sign occurs in Eq.
(\ref{magcur}) because of the convention in Fig. 1 where particle-number
current is positive for $V>0$.) \ In this article, the coefficient $\iota$ is
more convenient than the experimentally preferred low-voltage
tunneling-magnetoresistance ratio%
\begin{equation}
\text{TMR}=(R_{\text{AP}}-R_{\text{P}})/R_{\text{P}}=2\mathit{\iota
}/(1-\mathit{\iota}). \label{TMR}%
\end{equation}
\ \ 

The original equation due to Julliere \cite{Jul}, expressed in our notation by
the formula%
\begin{equation}
\mathit{\iota}=P_{\text{L}}P_{\text{R}},\label{J=PLPR}%
\end{equation}
enjoys considerable success in interpreting experiments \cite{MoodMatRev}.
\ We find below that whenever $\iota$ separates this way into two polarization
factors characteristic of the respective electrode-and-barrier compositions,
pseudo-torque expressions having dimensionless coefficients $\tau_{\text{L}}$
and $\tau_{\text{R}}$ [See Eqs. (\ref{TR=sin}), (\ref{TR=PL}), and
(\ref{TL=sin}) below.]$,$ whose simplicity parallels that of Eqs.(\ref{magcur}%
) and (\ref{J=PLPR}), hold also. \ The presence of the same average current
density $J_{0}(V)$ in equations both for magneto-current and torque represents
a strong connection between these two phenomena.

After the commonalities in Secs. 2 and 3, these mutual relations (Secs. 4 and
5) between magneto-conductance and pseudo-torques constitute the first of two
parts of the present article. \ The second part (Secs. 6 and 7) is stimulated
by the fact that theory does not \textit{generally} support the separability
of spin-channel currents into the left-dependent and right-dependent factors
needed to justify polarization factors in the first place. \ Previous theories
attack the question of polarization coefficients within the context of real
electron structure by considering the transmission of electrons initially
occupying well-defined crystalline-momentum states
\cite{TsymPet,MathUmer,Belash}. \ They posit either complete absence of
disorder or special types of disorder only within the barrier to legitimize
tunnel-polarization factors. \ The present approach, detailed below,
complements those works by \emph{excluding disorder only from a subregion of
the barrier.}%

\begin{figure}
[t]
\begin{center}
\includegraphics*[
width=3.039in
]%
{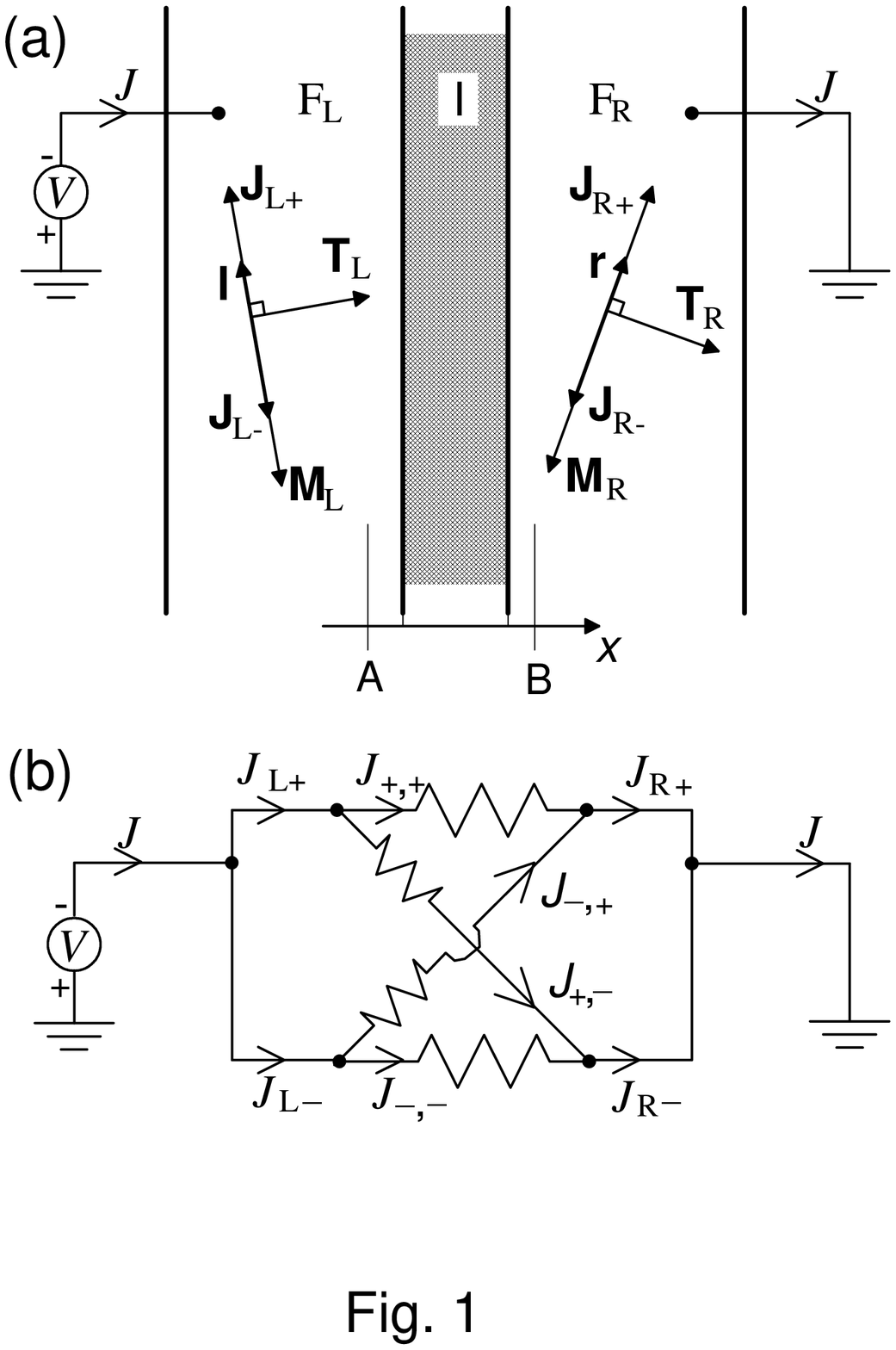}%
\caption{ (a) Scheme of magnetic tunnel junction and key to notations. (b)
Equivalent circuit for spin-channel currents and further key to notations.}%
\label{Fig1}%
\end{center}
\end{figure}

Electron scattering, which causes metallic resistivity, abounds within
experimental MTJ electrodes. \ A new feature of the present work is to forego
altogether crystal-momentum quantization within the electrodes. \ This feature
is particularly appropriate to contemporaneous experiments relying for
electrodes on evaporated or sputtered magnetic elements and alloys having high
defect concentration \cite{Nature AC,RipOsc,Nguyen,Fuchs}. Both alloying and
structural defects may cause an electron to scatter many times within the
electrodes before and after it tunnels across the barrier so that initial and
final crystal momenta are undefined.

Our elastic-tunneling theory rests on Bardeen's transfer-hamiltonian method
(BTM) \cite{Bard,Duke}, which is applicable to tunneling transitions between
thermal baths of electron states without any spatially conserved observables.
\ Bardeen defines two sets of basis states $-$ one for the left
electrode-and-barrier and one for the right electrode-and-barrier. \ Fermi's
"golden rule" for transition rates gives the tunneling current. \ Thus our
theory of MTJs has broader application than many others, previously reviewed
\cite{MoodMatRev,MTIRev,TsymbRev}, which rely on scattering of Bloch
electrons. \ Although more modern than Bardeen's method, they must assume
defined initial and final momenta.

Our model of the junction, called \textit{ideal middle, }excludes disorder
only from a central geometric slab of uniform thickness $w$, which may consist
of vacuum or periodic crystal lying somewhere within the barrier. \ We find
that exact factorization of channel-to-channel current, which leads to
Eqs.(\ref{J=PLPR}) and (\ref{TR=PL}) below,\ occurs in the limit
$w\rightarrow\infty,$ just as in the case of complete absence of disorder.
Further, our parametrization of lateral \textit{auto-corellation} (See Sec. 7)
of the Bardeen basis-function sets predicts well-defined
tunneling-polarization factors for finite barriers to first order in $w^{-1},$
which enhances their legitimacy for interpretation of experiments involving
any degree of disorder. \ Computations and measurements of the new
corellation-scale parameters $\xi_{\sigma}$ could shed quantitative light on
the genesis of polarization factors.

By way of organization, Section 1 is this Introduction and Section 2 shows how
spin-channel tunnel currents generally determine voltage-driven torque.
\ Section 3 uses the BTM to derive the resulting fully general expressions for
the magneto-conduction, the torques, and the relevant dimensionless
coefficients $\iota$, $\tau_{\text{L}},$ and $\tau_{\text{R}}$. \ Section 4
shows how tunneling-polarization factors and the resulting simple expressions
for $\iota$, $\tau_{\text{L}},$ and $\tau_{\text{R}}$ arise from a formal
separability condition. \ Section 5 addresses the expressions for
voltage-unsymmetric torque arising from voltage dependences of polarization
factors. \ Section 6 demonstrates the separation condition and derives the
tunnel-polarization factors which arise in the ideal-middle model at
$w\rightarrow\infty.$ \ Section 7 expands the magnetic tunneling properties
for finite $w$ and derives a formula for the first-order $w^{-1}$-dependence
of tunnel-polarization factors. \ Section 8 summarizes and discusses the results.

\section{First currents, then pseudo-torques}

Whenever two ferromagnets are separated by a nonmagnetic spacer, whether a
tunneling barrier or a metal, exchange-generated pseudo-torques acting on the
magnetic moments are attributable to the flow of spin-polarized current. \ For
a fuller discussion of the genesis of pseudo(or effective)-torque from the
principle of spin continuity, see Appendix B of Ref. \cite{'02}. \ Essential
is the interpretation of magnetization dynamics $(\overset{\mathbf{\bullet}%
}{\mathbf{M}}\equiv d\mathbf{M}/dt)$ governed by the additive terms in the
macroscopic Landau-Lifshitz equation. \ Ordinarily $\overset{\bullet
}{\mathbf{M}}$ represents the \textit{precession} in place of electron-spin
momentum localized to a volume element $dV$ due to local causes like magnetic
field, spin-orbit coupling, etc. \ But the term describing externally driven
spin transfer is transparently different. \ It reflects directly the
\textit{flow} of spin momentum \textit{directly} into $dV$. \ 

Indeed, the same may be said about the phenomenological exchange stiffness
described commonly by the effective field 2$A\nabla^{2}\mathbf{m,}$ with
$\mathbf{M\equiv}\mathit{M}_{\text{s}}\mathbf{m.}$ This truth is masked by the
derivability of ordinary exchange torque from variation of the stored energy
density $A\Sigma_{i,j}(\partial m_{i}/\partial x_{j})^{2}.$ Because spin
transfer is driven by an externally supplied current or voltage, its effect
cannot be derived from a stored energy. \ Therefore, its calculation requires
direct recourse to spin currents as detailed below. \ Since this distinction
between torque and divergence of polarization makes no difference in the
subsequent application of the Landau-Lifshitz equation, the prefix "pseudo-"
will be omitted in the remainder of this article. \ 

Return now to our problem of spin-transfer torque created by external voltage
applied to the MTJ. Consider particularly the series electric circuit in Fig.
1 (a) in which an external voltage $V$ causes electric-current density $J$ to
flow in series through a left metallic ferromagnetic film F$_{\text{L}}$, a
thin insulator I serving as a tunnel barrier, and finally a grounded right
metallic ferromagnetic film F$_{\text{R}}$. \ By assumption, F$_{\text{L}}$ is
sufficiently thin for the \textit{direction} of spontaneous magnetization
$\mathbf{M}_{\text{L}}(x)=$ $-M_{\text{L}}(x)\mathbf{l}$ within F$_{\text{L}}$
not to depend on the plane-perpendicular coordinate $x$;
similarly$\ \mathbf{M}_{\text{R}}(x)=$ $-M_{\text{R}}(x)\mathbf{r}$ within
F$_{\text{R}}.$ \ But the spontaneous magnetizations $M_{\text{L}}$and
$M_{\text{R}}$ may vary with $x$\textit{.} (Here the three-dimensional unit
vectors \textbf{l} and \textbf{r }include the angle $\theta=\cos
^{-1}\mathbf{r\cdot l.}$) \ Thus we lay aside those possibilities of forward
spin-wave excitation \cite{Berger96} and volume-intensive torque \cite{BJZ},
arising from dependence of magnetization direction on\textbf{\ }$x,$ which
become significant for larger film thickness and current density.

One goal is to calculate the component $T_{\text{R}}$ of interfacial torque
vector $\mathbf{T}_{\text{R}}$ per unit area, acting on $\mathbf{M}_{\text{R}%
},$ which lies orthogonal to \textbf{r} \textit{within the instantaneous
plane} common to $\mathbf{l}$ and \textbf{r} as indicated in Fig. 1a$.$ (The
orientation of the magnetic space spanned by \ $\mathbf{l}$ or \textbf{r} is
completely disconnected from that of position space $x,y,z.)$ \ A general
expression for $T_{\text{R}}$ \cite{'02,'99} reads thus:%
\begin{equation}
T_{\text{R}}=\hbar\lbrack J_{\text{L,}+}-J_{\text{L,}-}+(J_{\text{R,}%
-}-J_{\text{R,}+})\cos\theta]/2e\sin\theta\label{LRgen}%
\end{equation}
Here the left spin-channel electric current densities \textbf{J}%
$_{\text{L,}\pm}=J_{\text{L,}\pm}\mathbf{l}$ flow through plane A (See below)
in direction $x$ and the right \textbf{J}$_{\text{R,}\pm}=J_{\text{R,}\pm
}\mathbf{r}$ flow through plane B. \ The factor $-\hbar/2e$ converts any
electric channel current to one of spin momentum. \ A similar expression holds
for the pseudo-torque $T_{\text{L}}$ on the left moment. \ The torques
$T_{\text{R}}$ and $T_{\text{L}}$ must generally be included in the dynamic
Landau-Lifshitz equations for the two magnetic films.

Although previously applied only to all-metallic multilayers, Eq.
(\ref{LRgen}) may also be used when the spacer is an insulator. \ For its
derivation, one posits the non-relativistic n-electron hamiltonian including,
besides kinetic energy, coulomb terms due to external voltage and
electron-nuclear and electron-electron interactions. \ In addition, one
accepts the microscopically-based approximation, defensible in the case of Co,
that the transverse (to local \textbf{M}) components of conduction-electron
spin polarization created at the two internal I/F interfaces decay to zero
well within a characteristic distance $d_{\perp}\approx1$nm \cite{'02}, which
was estimated explicitly for Co/Cu and other interfacial compositions by
scattering computations \cite{SZ}. \ Moreover, in one experiment the threshold
current for switching of Co by polarized current flowing through a
\textit{metallic }spacer is simply proportional to film thickness down to 1
nm, confirming that the transverse polarization inside the ferromagnetic film
vanishes at this scale \cite{Cornthik}. \ Therefore the currents in the left
and right magnets must be polarized along instantaneous left (\textbf{l) }and
right (\textbf{r) }moment axes at depths greater than $d_{\perp}$ from the F/I
interfaces. \ Thus our work excludes thicknesses
$<$
1 nm, which require special treatment sensitive to atomic layering
\cite{Edwards Mathon}.

In the extensive literature on tunneling magnetoresistance involving Fe, Co,
Ni and magnetically concentrated alloys of these elements with others of lower
atomic number, there is little indication of spin relaxation at I/F
interfaces. \ Moreover experiments at cryogenic temperatures reveal that the
distance $\lambda_{||}$ of spin relaxation due to spin-orbit coupling for the
polarization component along the axis \textbf{M} is about 50 nm for Co and
about 5.5 nm for Ni-Fe \cite{BasSpRel}. \ Thus it follows that, at least in
the case of Co where $\lambda_{||}>>d_{\perp}$, the channel currents
$J_{\text{L,}\pm}$ and $J_{\text{R,}\pm}$\ should be evaluated at the planes A
and B lying at the\ distance $d_{\perp}$ from the respective F/I interfaces.
\ For within the space between these planes one may neglect spin-orbit effects
and embrace the well-known spin-continuity relation which equates the sum of
equivalent interfacial pseudo-torques with the net inflow of spin current
\cite{'02,'99}, having polarization directions \textbf{l} on the L side and
\textbf{r} on the R side. \ In the notation of Fig. 1 (a), the statistical
average of this equality becomes%
\begin{equation}
\mathbf{T}_{\text{L}}+\mathbf{T}_{\text{R}}=\frac{\hbar}{2e}\left[
(J_{\text{L,}-}-J_{\text{L,+}})\mathbf{l}+(J_{\text{R,}+}-J_{\text{R,}%
-})\mathbf{r}\right]  .\label{spincont}%
\end{equation}
By our assumed neglect of changes in $M_{\text{L}},$ we write $\mathbf{l\cdot
T}_{\text{L}}=0.$ \ Therefore the scalar product of Eq. (\ref{spincont}) with
\textbf{l }eliminates $\mathbf{T}_{\text{L}}$ and\textbf{\ }gives Eq.
(\ref{LRgen}) for the magnitude $T_{\text{R}}.$ \ A similar equation holds for
$T_{\text{L}}$.

The above argument neglects a \textit{decaying and spatially oscillating}
transverse current, calculated in certain FNF cases to lie between 0 and
$\simeq$10\% of the incident \ spin current (See Fig. 7 of Ref. \cite{SZ}). It
is likely due to specular interference created at the perfect interface
assumed in the calculation. \ Studies of FMF exchange coupling in vogue 10
years ago suggest that extremals in the Fermi surface determine the wavelength
and cause the amplitude to decay with distance. \ The amplitude will be
decreased by irregularities at real imperfect interfaces.

Even in the \textit{absence of applied electric voltage }($V=0)$\textit{\ }an
additional \textit{perpendicular} component of exchange pseudo-torque
$\mathbf{T}_{\text{R}\perp}=K\mathbf{l\times r=-\mathbf{T}_{\text{L}\perp}}$
predicted for MTJs \cite{'89} is generally related to phenomenological
coupling energy $-K\mathbf{l\bullet r=-}K\cos\theta$. \ It must also be
included in the Landau-Lifshitz equation for the dynamics of magnet
F$_{\text{R. \ }}$However, in that toy rectangular-barrier MTJ model
\cite{'89}, the (uncalculated) dependence of $\mathbf{T}_{\text{R}\perp}$ and
$\mathbf{\mathbf{T}_{\text{L}\perp}}$ on applied voltage occurred only in
higher order $(\varpropto V^{2})$ than the torque given by Eq. (\ref{LRgen})
$(\varpropto V)$. \ Moreover, its dynamic effect is relatively weaker in
structures with coincident easy anisotropy axes and low magnetic damping, such
as the pillars using metallic spacers experimentally favored for efficient
current-driven switching \cite{Nature AC}. \ Indeed, steady oscillation
excited by a steady electric current, such as that observed \cite{Nature
AC,RipOsc}, is possible with $\mathbf{T}_{\text{R}\perp}=0,$ but not in the
absence of in-plane $\mathbf{T}_{\text{R}}.$ In addition, the BTM used here
does not readily provide this out-of-plane torque. \ For these reasons, we do
not attempt to predict the perpendicular torque component in this work.

\section{Magneto-conduction and torques}

Equation (\ref{LRgen}) effectively reduces the interacting-electron problem of
voltage-driven torque to the customarily independent-electron problem of
spin-channel currents. \ One recently reviewed BTM-based theory of collinear
MTJ magnetoresistance \cite{MTIRev} extends naturally to tunneling between
spin channels for general $\theta.$ \ For adaptation of the BTM
\cite{Bard,Duke} to the MTJ of Fig. 1a, a stationary basis state
$|p,\sigma\rangle$ within the electron reservoir F$_{\text{L}}$ is assigned
orbital index \emph{p} and majority/minority spin $\sigma=\pm$ quantized along
axis \textbf{l}. \ It satisfies $(H+eV-\epsilon_{p,\sigma})|p,\sigma
\rangle=0,$ and decays exponentially within the barrier, considered
semi-infinite in width when defining the basis states. \ Here, $H=p^{2}%
/2m+\Sigma_{\sigma}|\sigma\rangle U_{\sigma}(x,y,z)\langle\sigma|,$ where the
potential $U_{\sigma}$ depends on spin within the ferromagnets according to
intinerant-electron magnetism theory \cite{Kubler}, but not within the
barrier. \ Within F$_{\text{R}},$ a\ similar state satisfies $(H-\epsilon
_{q,\sigma^{\prime}})|q,\sigma^{\prime}\rangle=0$ \ with quantization axis
\textbf{r}. \ Because the barrier is assumed to dominate all other resistances
of this circuit, the spin channels are shown in Fig. 1 (b) as shorted in each
magnet and/or external-contact region by spin lattice relaxation due to
spin-orbit coupling. \ One may disregard \textit{spin accumulation} and the
related distinction between electric and electrochemical potentials which are
important when a non-magnetic metallic spacer substitutes for the barrier
\cite{ValetFert}. \ $U_{\sigma}$ includes all elastic terms arising from
atomic disorder due to alloying, defects, interfacial atomic interdiffusion,
etc. \ The state indices $p,q$ simply enumerate the exact eigenstates
$|p,\sigma\rangle,|q,\sigma^{\prime}\rangle$ of $H$ in the Bardeen basis.
\ Each such state incorporates effects of all multiple elastic scatterings
without limit.

Employing the spinor transformation connecting quantization axes \textbf{l
}and \textbf{r}, the transfer matrix element takes the form%
\begin{equation}
\langle p,\sigma|H-\varepsilon|q,\sigma^{\prime}\rangle=%
\begin{bmatrix}
\gamma_{p,+;q,+}\cos\frac{\theta}{2} & \gamma_{p,+;q,-}\sin\frac{\theta}{2}\\
-\gamma_{p,-;q,+}\sin\frac{\theta}{2} & \gamma_{p,-;q,-}\cos\frac{\theta}{2}%
\end{bmatrix}
. \label{tunmat}%
\end{equation}
\ Direct extension of BTM \cite{Duke18.34} to our spin-dependent case gives
the expression%
\begin{equation}
\gamma_{p,\sigma;q,\sigma^{\prime}}(x)=\frac{-\hbar^{2}}{2m}%
{\textstyle\int}
dydz(\psi_{p,\sigma}\partial_{x}\varphi_{q,\sigma^{\prime}}-\varphi
_{q,\sigma^{\prime}}\partial_{x}\psi_{p,\sigma}), \label{orbmat}%
\end{equation}
where the integral is over unit area for coordinate $x$ lying appropriately
(see below) inside the barrier. \ The energies $\epsilon_{p,\sigma}\ $and
$\epsilon_{q,\sigma^{\prime}}$ may differ only infinitesimally from the Fermi
value $\varepsilon=\varepsilon_{\text{F}}.$ \ The hamiltonian $H,$ the left
($\psi_{p,\sigma})$ and right $(\varphi_{q,\sigma^{\prime}})$ orbital wave
functions, and these matrix elements (\ref{orbmat}) are real.

Only the neglect of cross-barrier overlaps $\langle p,\sigma|q,\sigma^{\prime
}\rangle$ allows use of the Fermi golden rule of perturbation theory which is
strictly valid for an orthonormal basis. Substitution of the perturbation
(\ref{tunmat}) into this rule is followed by summation over the initial states
in an infinitessimal energy band of width $eV.$ Thus the partial electric
current density flowing between channel $\sigma$ in \ F$_{\text{L}}$ and
channel $\sigma^{\prime}$ in F$_{\text{R}}$ becomes%

\begin{equation}
J_{\sigma,\sigma^{\prime}}=\frac{-2\pi e^{2}V}{\hbar}%
{\textstyle\sum\nolimits_{p,q}^{^{\prime}}}
\langle p,\sigma|H-\varepsilon_{\text{F}}|q,\sigma^{\prime}\rangle
^{2}\label{tuncur}%
\end{equation}
at $T=0$ K. \ The $^{\prime}$ in $%
{\textstyle\sum\nolimits_{p,q}^{^{\prime}}}
$ imposes the conditions $\varepsilon_{\text{F}}<(\varepsilon_{p,\sigma
},\varepsilon_{q,\sigma^{\prime}})<\varepsilon_{\text{F}}+eV.$

Notations in the equivalent circuit shown in Fig. 1 (b) make plain the
relations%
\begin{equation}
J_{\text{L}\sigma}=J_{\sigma,+}+J_{\sigma,-}\text{ },\text{ \ \ \ }%
J_{\text{R}\sigma^{\prime}}=J_{+,\sigma^{\prime}}+J_{-,\sigma^{\prime}},\text{
\ \ }(\sigma,\sigma^{\prime}=\pm)\label{circt}%
\end{equation}
needed in Eq. (\ref{LRgen}). \ The right hand sides of these equations are
evaluated from Eqs. (\ref{tunmat}-\ref{tuncur}).

Next we write the total electric current density $J=J_{\text{L,}%
+}+J_{\text{L,}-}.$ With the notation%

\begin{equation}
\Gamma_{\sigma,\sigma^{\prime}}=\frac{2\pi eV}{\hbar}%
{\textstyle\sum\nolimits_{p,q}^{^{\prime}}}
\gamma_{p,\sigma;q,\sigma^{\prime}}^{2} \label{Gamsum}%
\end{equation}
for interchannel particle-number tunneling conduction with the angular factor
omitted, the above equations\ combine to give Eq. (\ref{magcur}) with%
\begin{equation}
J_{0}=e(\Gamma_{+,+}+\Gamma_{-,-}+\Gamma_{+,-}+\Gamma_{-,+})/2 \label{J0}%
\end{equation}
and the electric magneto-conduction coefficient%
\begin{equation}
\mathcal{\iota}=e(\Gamma_{+,+}+\Gamma_{-,-}-\Gamma_{+,-}-\Gamma_{-,+})/2J_{0}.
\label{Jgen}%
\end{equation}
Eq.(\ref{LRgen}) becomes%
\begin{equation}
T_{\text{R}}=-(\hslash\tau_{\text{R}}J_{0}/2e)\sin\theta\label{TR=sin}%
\end{equation}
or, in coordinate-free form%
\begin{equation}
\mathbf{T}_{\text{R}}=(\hslash\tau_{\text{R}}J_{0}/2e)\mathbf{r\times(l\times
r),} \label{TRvec}%
\end{equation}
with the torque coefficient
\begin{equation}
\tau_{\text{R}}=e(\Gamma_{+,+}+\Gamma_{+,-}-\Gamma_{-,-}-\Gamma_{-,+})/2J_{0}.
\label{TRgen}%
\end{equation}
The fact that the linear combination of the parameters $\Gamma_{\sigma
,\sigma^{\prime}}$ appearing in Eq. (\ref{Jgen}) differs from that in Eq.
(\ref{TRgen}) and a similar one for $\mathbf{T}_{\text{L}}$ precludes any
fully general connection between torques and electrical current.

\section{Left-right separability and polarization factors}

Particularly interesting relations arise if the summation in Eq.
(\ref{Gamsum}) for the inter-channel particle current happens to separate into
left- and right-dependent factors in the form%
\begin{equation}
\Gamma_{\sigma,\sigma^{\prime}}=f\Omega_{\text{L},\sigma}\Omega_{\text{R,}%
\sigma^{\prime}}.\label{factoring}%
\end{equation}
Here the coefficient $f$ , which we make no attempt to evaluate, is
independent of $\sigma,\sigma^{\prime}$. \ (Sections 6 and 7 address
conditions for this separability.) \ For then Eq. (\ref{J0}) gives%
\begin{equation}
J_{0}=\frac{ef}{2}(\Omega_{\text{L},+}+\Omega_{\text{L},-})(\Omega
_{\text{R},+}+\Omega_{\text{R},-})\label{meanj}%
\end{equation}
and Eq. (\ref{Jgen}) gives Eq. (\ref{J=PLPR}) with the \textit{tunneling
polarization} parameters
\begin{equation}
\text{\ }P_{i}=\frac{\Omega_{i,+}-\Omega_{i,-}}{\Omega_{i,+}+\Omega_{i,-}%
}\text{ \ \ \ (}i=\text{L,R)}\label{polfac}%
\end{equation}
which are directly measurable using FIS junctions \cite{MoodMatRev}. \ In
these terms, Eqs. (\ref{magcur})\ and (\ref{J=PLPR}) give the
magneto-conduction and Eq. (\ref{TRvec}) the torque with
\begin{equation}
\tau_{\text{R}}=P_{\text{L}}.\label{TR=PL}%
\end{equation}
Similarly, the torque on the left magnet is%
\begin{equation}
T_{\text{L}}=-(\hslash\tau_{\text{L}}J_{0}/2e)\sin\theta\mathbf{,}\text{
\ }\tau_{\text{L}}=P_{\text{R}}\label{TL=sin}%
\end{equation}
or, in coordinate-free form%
\begin{equation}
\mathbf{T}_{\text{L}}=\frac{\hbar\tau_{\text{L}}}{2e}J_{0}\mathbf{l\times
(r\times l)}.\label{TL=PR}%
\end{equation}
\ The Eqs.(\ref{J=PLPR}), (\ref{TR=PL}), and (\ref{TL=PR}) show the very close
relation between current-driven torques and magneto-conduction at the same
voltage, summarized by $\iota=\tau_{\text{L}}\tau_{\text{R}},$ if the
separability condition (\ref{factoring}) is satisfied$.$

The ground-breaking paper of Julliere \cite{Jul} gave equations equivalent to
(\ref{J=PLPR}) and (\ref{polfac}) taking $\Omega_{\text{L}\sigma}$ and
$\Omega_{\text{R}\sigma^{\prime}}$ to be spin-dependent basis-state densities
at $\varepsilon=\varepsilon_{\text{F}}$. \ It appeared to attribute the
dimensionless magneto-current coefficient $\iota=P_{\text{L}}P_{\text{R}}$ to
bulk properties of the two magnetic compositions involved. \ But the
analytically solved free-electron rectangular-potential model \cite{'89} shows
that an interface-dependent factor must be included in $\Omega_{i,\sigma}$ as
well. \ The transfer-hamiltonian treatment of this toy model follows
immediately from the spinless treatment \cite{Duke18.34} giving
\begin{equation}
\Omega_{i,\sigma}=k_{i,\sigma}/(\kappa_{0}^{2}+k_{i,\sigma}^{2})\label{rectP}%
\end{equation}
where%
\begin{equation}
k_{i,\sigma}^{2}=2mE_{i,\sigma}/\hslash^{2}\text{ \ and \ }\kappa_{0}%
^{2}=2mB/\hslash^{2}.\label{k and kap}%
\end{equation}
Here, $E_{i,\sigma}$ is the kinetic energy at the Fermi level and $B$ is the
barrier potential measured from the Fermi level. \ Equation (\ref{polfac}) now
gives%
\begin{equation}
P_{i}=\frac{k_{i,+}-k_{i,-}}{k_{i,+}+k_{i,-}}\cdot\frac{\kappa_{0}^{2}%
-k_{i,+}k_{i,-}}{\kappa_{0}^{2}+k_{i,+}k_{i,-}}\label{polfree}%
\end{equation}
in agreement with Ref. \cite{'89}. \ In this formula, the first factor depends
purely on basis-state densities in the magnet, while the second mixes magnet
and barrier properties. \ The results of the toy model \cite{'89} satisfy the
general magneto-conduction relations (\ref{magcur}), (\ref{J=PLPR}) and torque
relations (\ref{TRvec}),(\ref{TR=PL}),(\ref{TL=PR}) with this substitution.

We note in passing that experimental variation of barrier height $B$ shows
considerable support for the zero of $\iota$ at the barrier potential
satisfying $\kappa_{0}^{2}-k_{i,+}k_{i,-}=0$ expected from Eq. (\ref{polfree})
\cite{MiyazZero} (for small $V$)$.$ Therefore, in spite of its fundamental
naivete, this toy model enjoys some degree of credibility. \ It illustrates
the general fact that, even when separability holds, each polarization factor
is a property of the electron structure of the magnet and barrier
\textit{combination }as demonstrated by many experiments and
calculations\textit{.} \ Section 7 will discuss how tunnel polarization may
vary with barrier thickness.

\section{Finite bias and torque asymmetry}

In experiments, TMR typically decreases significantly with increasing finite
$V$ \cite{MoodMatRev}. Voltage-dependence of interfacial transmission, special
state density distributions, extrinsic impurity effects, and inelastic
tunneling contribute to this decrease \cite{MoodMatRev,TsymbRev}. \ This is
important because large voltages will be required to read and write in a
2-terminal memory element.

The toy polarizations of Eq. (\ref{polfree}) will serve to illustrate
qualitatively the very unsymmetric effect of finite $V$ on voltage-driven
pseudo-torque. \ One calculation of TMR uses the WKB approximation for the
free-electron wave function within the constant-slope barrier potential
sketched in Fig. 2 \cite{Nanjing}. \ The interfacial transmissions are
approximated by those of the flat-potential polarizations (\ref{polfree}).
\ The authors cite some experimental support for their results.

It is the decrease of $P_{i}$ in the particular electrode which
\textit{collects }the tunneled electrons that primarily accounts for the
decrease of $\iota$ in the calculated result \cite{Nanjing}. \ In Fig. 2, for
$V>0,$ the collecting electrode lies on the right. \ Note that the electrons
whose energy lies in a narrow band (shaded in Fig. 2) just below the Fermi
level of the emitting electrode on the left of the barrier dominate the
tunneling current because of the strong energy dependence of the WKB factor
$\exp[-2\int\kappa(x)dx]$ in the transmission coefficient. \ Since these hot
electrons lie an amount well above the Fermi level on the right, this energy
shift $eV$ must be taken into account when estimating $P_{\text{R}}.$%

\begin{figure}
[th]
\begin{center}
\includegraphics*[
width=4.0421in
]%
{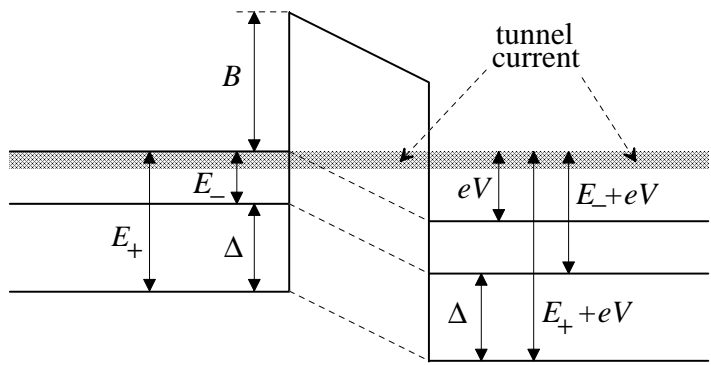}%
\caption{ \ Schematic junction potential for finite V. The shaded bar
indicates the energy range of most of the tunneling electrons.}%
\end{center}
\end{figure}

We simplify this model one step further and neglect the width of the shaded
current band in Fig. 2. It is then clear that Eqs. (\ref{k and kap}) and
(\ref{polfree}) with $i=$L\ are still correct for $P_{\text{L}},$ neglecting
correction for the finite slope of the barrier potential$.$ \ However, the equations%

\begin{equation}
k_{\text{R}\sigma}^{2}=2m(E_{\text{R}\sigma}+eV)/\hslash^{2}\text{ \ and
\ }\kappa_{0}^{2}=2m(B-eV)/\hslash^{2},\label{Vshift}%
\end{equation}
obtained by adding $eV$ to each electron energy on the right, must replace
Eqs.(\ref{k and kap}) for $i=$R.

Figure 3 plots the curves $\tau_{\text{L}}=P_{\text{R}}$ and $\tau_{\text{R}%
}=P_{\text{L}}$ evaluated from the preceding three equations as well as
TMR$\ $from Eqs. (\ref{TMR}) and (\ref{J=PLPR}) versus $V$ for the special
example of a symmetric junction with the parameters $k_{\text{L}-}%
=k_{\text{R}-}\equiv k_{-},$ $k_{\text{L}+}=k_{\text{R}+}\equiv10k_{-},$ and
$\kappa_{0}=6.4k_{-,}$\ whereby each electrode has the $V=0$ polarization
$P_{\text{L}}$=$P_{\text{R}}=0.5.$ In this illustration, TMR$(V)$ is symmetric
because it involves both $P_{\text{L}}$ and $P_{\text{R}}$ but $P_{\text{L,R}%
}(V)$ and the torque coefficients $\tau_{\text{L,R}}(V)$ are not. \ Although
the theory preceding this section assumed small $V,$ the present discussion
makes reasonable the application of the results to finite $V$ with the
understanding that the polarization of the collecting electrode generally
depends more strongly on $V.$ \ Of course, this toy calculation cannot make
quantitative predictions of the $V$-dependence which must rest on details of
electron structure \cite{MoodMatRev,TsymbRev}.%

\begin{figure}
[th]
\begin{center}
\includegraphics*[
width=4.0421in
]%
{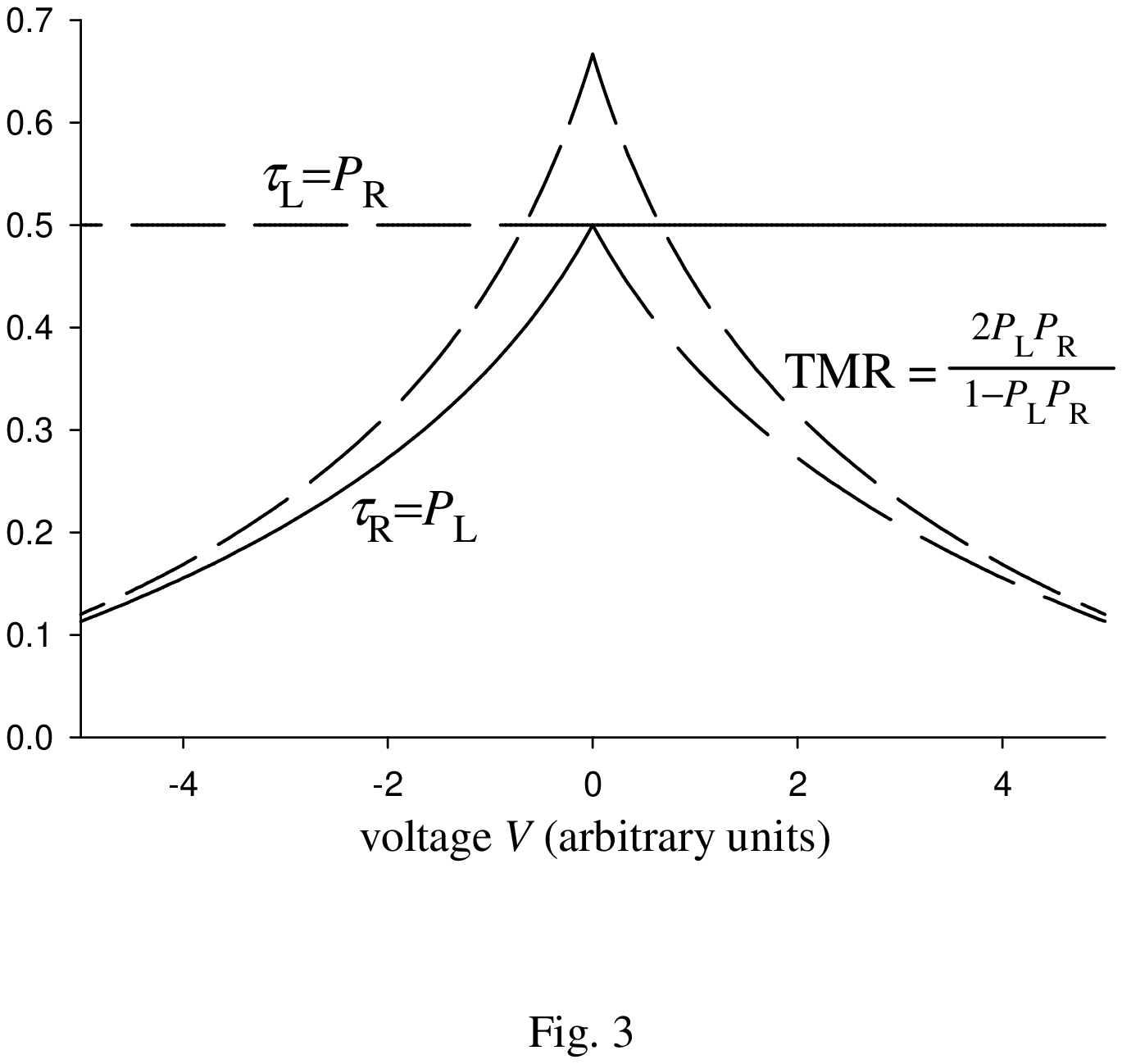}%
\caption{ \ Schematic effect of finite voltage on TMR, polarization, and
torque coefficients illustrated by the toy free electron model of a physically
symmetric magnetic tunnel junction. \ Note that TMR is symmetric but the other
coefficients are not. \ The parameters are $\kappa_{0}=6.4k_{-}$,
$k_{+}=10k_{-}.$}%
\label{Fig3}%
\end{center}
\end{figure}

Note that while critical \textit{current density }for magnetic
excitation\textit{\ }is appropriate to junctions with metallic spacers, the
high resistance of a MTJ makes critical voltage more appropriate. \ (Indeed,
strictly speaking, the critical current of a \textit{constant-current}
generator will generally \textit{differ} from the current density flowing at
threshold in the presence of constant external voltage.) \ Another significant
difference between metallic and insulating spacers lies in the angular
symmetry of the torque. \ The fixed $\sin\theta$-dependence at constant $V$ in
the tunneling case has no counterpart in the metallic case where more general
torque expressions typically contribute to asymmetry of excitation threshold
\cite{'02}. \ Now we see that the non-ohmic resistance of a tunneling barrier
gives rise to the torque asymmetry of $\tau_{\text{R}}(V)$ exhibited in Fig.
3, which naturally reflects in yet another origin for asymmetry of voltage threshold.

\section{Ideal-middle model for separability}

A recent publication compares existing theoretical arguments supporting the
existence of tunnel-polarization factors \cite{Belash}. \ Each of them assumes
incident states with definite crystalline momentum. One common type of
argument assumes complete absence of disorder so that the tunneling through a
thick barrier is dominated by a single value of lateral momentum. \ A
different model of Tsymbal and Pettifor \cite{TsymPet} recovers factorization
and therefore the Julliere formula in a tight-binding single-band model
\textit{disordered only within the barrier}.\ \ Similarly, the model of Mathon
and Umerski attributes the factorization to phase decoherence due to disorder
within the barrier \cite{MoodMatRev,MathUmer}. \ These treatments are
augmented with arguments based on the Feynman path integral in a disordered
barrier \cite{Belash}. \ Our treatment below complements these arguments with
the contrary tack of foregoing lateral momentum quantization completely within
the electrodes and I/F interfaces while preserving ideal crystalline ordering
or vacuum within the middle of the barrier.%

\begin{figure}
[th]
\begin{center}
\includegraphics*[
trim=0.000000in 0.000000in 0.000000in 0.669642in,
width=3.013in
]%
{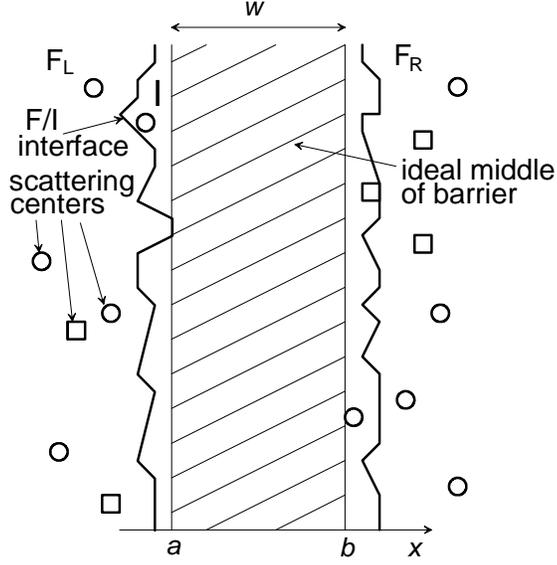}%
\caption{ \ Depiction of the ideal-middle model of a magnetic tunneling
junction. \ Disorder without limit is permitted in both electrodes and barrier
except within a central slab $\mathcal{B}$ of the barrier lying between the
portal planes $x=a,b.$}%
\label{Fig4}%
\end{center}
\end{figure}
\ Figure 4 indicates the structural scheme. The left ($\psi_{p,\sigma})$ and
right $(\varphi_{q,\sigma^{\prime}})$ orbital basis functions for the transfer
matrix, introduced in Sec. 3, are governed in detail by the general potential
$U_{\sigma\text{ or }\sigma^{\prime}}$ depending on crystal structure, alloy
composition, defects, F/I interface roughness and atomic interdiffusion, etc.
\ The quantum numbers $p$ and $q$ do not refer to any diagonal operator.
\ Exceptionally, the \textit{ideal-middle} $\mathcal{B}$ of the barrier
consists of an ideal crystalline slab or vacuum region defined by $a\leq x\leq
b$ where the planes $x\mathit{=}a,b$ are dubbed \textit{portals} of the ideal
middle. \ In order to define the left and right basis-state sets of the
Bardeen theory, the barrier potential extends into alternative semi-infinite
spaces $(a\leq x)$ and ($x\leq b)$\ , where it is greater than $\varepsilon
_{\text{F}},$ independent of or periodically dependent on $y,z$ and
independent of $\sigma$ and $\sigma^{\prime}.$ The respective conditions
$\psi_{p,\sigma}\rightarrow0$ for $x\rightarrow\infty$ and $\varphi
_{q,\sigma^{\prime}}\rightarrow0$\ for $x\rightarrow-\infty$ complete the
definitions of $\psi_{p,\sigma}$ and $\varphi_{q,\sigma^{\prime}}$.

The effective-mass theorem \cite{SJA} is valid when $\varepsilon$ is near the
bottom $\mathbf{k=k}_{0}$ of the conduction band within region $\mathcal{B}$.
\ Then the evanescent portion of a left-magnet basis function within this
region is approximated by%
\begin{equation}
\psi_{p,\sigma}=\Psi_{p,\sigma}(x,y,z)u_{\text{cb,}\mathbf{k}_{0}%
}(x,y,z)\label{evanwav}%
\end{equation}
where $\Psi_{p,\sigma}$ satisfies ($\mathcal{H}_{\text{bar}}-\varepsilon
_{p,\sigma})\Psi_{p,\sigma}\mathcal{=}0$ and $\Psi_{p,\sigma}\rightarrow0$ for
$x\rightarrow\infty,$ and $u_{\text{cb,}\mathbf{k}_{0}}$ is the Bloch function
for the bottom of the conduction band. \ The effective barrier hamiltonian is
$\mathcal{H}_{\text{bar}}=-\hbar^{2}\nabla^{2}/2m_{\text{cb}}+\mathcal{U}(x)$
where $m_{\text{cb}}$ is the effective mass and $\mathcal{U(}x)$
$(>\varepsilon_{\text{F}})$ is the spin-independent atomically smoothed
effective barrier potential. \ Similarly for F$_{\text{R}},$ $\varphi
_{q,\sigma^{\prime}}=\Phi_{q,\sigma^{\prime}}u_{\text{cb,}\mathbf{k}_{0}}$
with $\Phi_{q,\sigma^{\prime}}\rightarrow0$ for $x\rightarrow-\infty.$ \ In
case of vacuum, ($\Psi,\Phi)$ are indistinguishable from ($\psi,\varphi).$
\ (Note however that this treatment fails if both\textit{ }$V$ is
finite\textit{ and\ }the FI\ interfaces are disordered,\textit{ }for then
$\mathcal{U}$ depends on $y$ and $z$ as well as $x$.)

Assuming periodic boundary conditions in the \textbf{s}$\mathbf{=(}y,z)$
sub-space, the evanescent portions of left and right basis states within
$\mathcal{B}$ are conveniently fourier-expanded in space \textbf{s} with the
WKB approximation giving
\begin{equation}
\Psi_{p,\sigma}=\sum\nolimits_{\text{\textbf{k}}}\lambda_{p,\sigma}%
(\mathbf{k})[\kappa(k,a)/\kappa(k,x)]^{1/2}\exp\left[  -\int_{a}^{x}%
\kappa(k,x^{\prime})dx^{\prime}+i\mathbf{k}\bullet\mathbf{s}\right]
\label{lpsi}%
\end{equation}
and%
\begin{equation}
\Phi_{q,\sigma^{\prime}}=\sum\nolimits_{\text{\textbf{k}}}\mu_{q,\sigma
^{\prime}}(\mathbf{k})[\kappa(k,b)/\kappa(k,x)]^{1/2}\exp\left[  -\int_{x}%
^{b}\kappa(k,x^{\prime})dx^{\prime}+i\mathbf{k}\bullet\mathbf{s}\right]
\label{rphi}%
\end{equation}
where the sums $\Sigma_{\text{\textbf{k}}}$ are carried over a 2-dimensional
reduced Brillouin zone. \ These formulas employ the function%

\begin{equation}
\kappa(k,x)\emph{\ }=\left[  \kappa_{0}^{2}(x)+k^{2}\right]  ^{1/2},\text{
\ \ \ with\ }\kappa_{0}^{2}=2m_{\text{cb}}[\mathcal{U}\left(  x\right)
-\varepsilon_{\text{F}}]/\hbar^{2},\label{kap}%
\end{equation}
where $i\kappa$ is the imaginary component of the wave-vector in region
$\mathcal{B}$.\ \ Note that Eqs. (\ref{lpsi}) and (\ref{rphi}) reduce to
expansions of $\psi_{p,\sigma}$ and $\varphi_{q,\sigma^{\prime}}$ with
coefficients $\lambda_{p,\sigma}(\mathbf{k})$ and $\mu_{q,\sigma^{\prime}%
}(\mathbf{k})$ on the portal planes $x=a$ and $x=b$ respectively.

The transfer-hamiltonian matrix element of Eq. (\ref{orbmat}) is evaluated at
any $x$ lying within the interval $a\leq x\leq b$. \ Consequently $\Psi,$
$\Phi$ , and \textit{m}$_{\text{cb}}$ may replace $\psi$, $\varphi,$ and
\textit{m} respectively in this formula. \ One convenient choice to evaluate
Eq. (\ref{orbmat}) is $x=x_{\text{max}},$ satisfying $\mathcal{U(}x)\leq$
$U(x_{\text{max}})$ for all $x,$ because the resulting condition
$\partial\kappa_{0}/\partial x(x_{\text{max}})=0$ simplifies the mathematics.
\ (Inclusion in $\mathcal{U}$ of the image potential due to electron-electron
correlation will often insure the presence of a maximum, even if $|V|$ is
large.) \ Substitution of Eqs. (\ref{lpsi}) and(\ref{rphi}) followed by
integration over $y$ and $z$, with the assistance of the identity $\int
d$\textbf{s}$^{2}\exp[i(\mathbf{k}-\mathbf{k}^{\prime})\cdot$\textbf{s}%
$\mathbf{]=\delta}_{\mathbf{k},\mathbf{k}^{\prime}},$ reduces Eq.
(\ref{orbmat}) to%
\begin{equation}
\gamma_{p,\sigma;q,\sigma^{\prime}}=\Sigma_{\mathbf{k}}F(w,\mathbf{k)}%
\lambda_{p,\sigma}^{\ast}(\mathbf{k})\mu_{q,\sigma^{\prime}}(\mathbf{k}%
)\label{nugam}%
\end{equation}
where%
\begin{equation}
F(w,\mathbf{k)=}\frac{-4\pi^{2}\hbar^{2}}{m_{\text{cb}}}\kappa^{1/2}%
(\mathbf{k},a)\kappa^{1/2}(\mathbf{k},b)\exp[-\int_{a}^{b}dx\kappa
(k,x)].\label{F(k)}%
\end{equation}
Here we use the barrier-middle thickness $w=b-a,$ and note $\lambda_{p,\sigma
}^{\ast}(\mathbf{k})=\lambda_{p,\sigma}(-\mathbf{k})$ and $\mu_{q,\sigma
^{\prime}}^{\ast}(\mathbf{k})=\mu_{q,\sigma^{\prime}}(-\mathbf{k})$ because
$\Psi_{p,\sigma}$ and $\Phi_{q,\sigma^{\prime}}$ are real. \ [When $w$ varies
in our discussion below, $\lambda_{p,\sigma}(\mathbf{k})$ and $\mu
_{q,\sigma^{\prime}}(\mathbf{k})$ remain unchanged because they pertain to the
semi-infinite barrier independent of $w.$ \ We merely expand or contract the
ideal middle of the barrier in Eq. (\ref{F(k)}).] \ After rearranging the
order of sums, Eq. (\ref{Gamsum}) with substitution of (\ref{nugam}) and
(\ref{F(k)}) becomes%
\begin{equation}
\Gamma_{\sigma,\sigma^{\prime}}=\frac{2\pi eV}{\hslash}\sum
\nolimits_{\text{\textbf{k}}}F(w,\mathbf{k})\sum\nolimits_{\text{\textbf{k}%
}^{\prime}}F(w,\mathbf{k}^{\prime})\mathcal{L}_{\sigma}(\mathbf{k,k}^{\prime
})\mathcal{M}_{\sigma^{\prime}}(\mathbf{k,k}^{\prime})\label{Gam}%
\end{equation}
where each of the two functions
\begin{equation}
\mathcal{L}_{\sigma}=\Sigma_{p}^{^{\prime}}\lambda_{p,\sigma}^{\ast
}(\mathbf{k})\lambda_{p,\sigma}(\mathbf{k}^{\prime}),\text{ }\mathcal{M}%
_{\sigma^{\prime}}=\Sigma_{q}^{^{\prime}}\mu_{q,\sigma^{\prime}}%
(\mathbf{k})\mu_{q,\sigma^{\prime}}^{\ast}(\mathbf{k}^{\prime})\label{LM}%
\end{equation}
depends only on parameters of the left and right magnet-and-barrier
combinations respectively. \ The $^{\prime}$ on $\Sigma^{\prime}$ signifies
the conditions given previously for Eq. (\ref{tuncur}).

In the presence of atomic disorder, the sums in Eqs. (\ref{LM}) are carried
over many states of randomized character. Therefore they have the nature of
statistical auto-correlations in $y,z$-space which should depend smoothly on
\textbf{k }and $\mathbf{k}^{\prime}$ and are Taylor-expandable about
$\mathbf{k=k}^{\prime}=0$. \ (See Sec. 5 for the very different toy
free-electron case of vanishing disorder \cite{'89}, in which one may formally
replace $p\rightarrow$\textbf{k}$^{\prime\prime},$ $q\rightarrow$%
\textbf{k}$^{\prime\prime\prime}$\ so that $\mathcal{L}_{\sigma}$ and
$\mathcal{M}_{\sigma^{\prime}}$ become proportional to $\delta_{\mathbf{k,k}%
^{\prime}}$.) \ In addition, with increasing thickness $w=b-a$ of region
$\mathcal{B}$, the exponential in Eq.(\ref{F(k)}) becomes ever more sharply
peaked at $\mathbf{k}=0.$ \ Summation over $\mathbf{k}$ and $\mathbf{k}%
^{\prime}$ of the terms in these Taylor series' for finite $w$ gives the
corresponding terms%
\begin{equation}
\Gamma_{\sigma,\sigma^{\prime}}(w)=\Gamma_{\sigma,\sigma^{\prime}}%
^{(0)}(w)+\Gamma_{\sigma,\sigma^{\prime}}^{(1)}(w)+\text{ }\cdot\cdot
\cdot\label{Gamser}%
\end{equation}
The initial constants in both Taylor expansions yield%
\begin{equation}
\Gamma_{\sigma,\sigma^{\prime}}^{(0)}(w)=f(w)\Omega_{\text{L,}\sigma}%
^{(0)}\Omega_{\text{R,}\sigma}^{(0)}\label{G0fact}%
\end{equation}
with $\Omega_{\text{L,}\sigma}^{(0)}\equiv\mathcal{L}_{\sigma}(0,0)$
and\ \ $\Omega_{\text{R,}\sigma^{\prime}}^{(0)}\equiv\mathcal{M}%
_{\sigma^{\prime}}(0,0).$ \ Here factors independent of $\sigma$ and
$\sigma^{\prime}$ are absorbed into $f.$ \ Therefore, to leading order in this
expansion, the integrations in Eq. (\ref{Gam}) tend to the left-right
separation of the form (\ref{factoring}).

Written in full, the parameters needed in the general polarization formula
(\ref{polfac}) are, to lowest order in the Taylor expansions of Eqs. (33), the
basis-\textit{state weights}%
\begin{align}
\Omega_{\text{L},\sigma}^{(0)} &  =\sum_{p}\left(  \int\int dydz\Psi
_{p,\sigma}(a,y,z)\right)  ^{2}\label{OmegL}\\
\Omega_{\text{R},\sigma^{\prime}}^{(0)} &  =\sum_{q}\left(  \int\int
dydz\Phi_{q,\sigma^{\prime}}(b,y,z)\right)  ^{2}\label{OmegR}%
\end{align}
where $\int\int dydz$ \ is carried over unit junction area at the portal
positions $a$ and $b.$ \ [See the next section for development of
$\Gamma_{\sigma,\sigma^{\prime}}^{(1)}(w).]$ \ \textit{Note that the latter
two equations differ generally from the} \textit{local state (or charge)
density} often cited in connection with tunneling. \ (LSD $\varpropto\int\int
dydz\Psi_{p,\sigma}^{2})$ \ They reduce to the LSD in the complete absence of
disorder when each of the two sums reduces to a single term $\Psi
_{\mathbf{k}=0,\sigma}^{2}$ and $\Phi_{\mathbf{k}=0,\sigma^{\prime}}^{2}$
independent of \textit{y} and \textit{z}.

\section{Correction of polarization at finite thickness}

The non-orthogonality between left and right basis functions constitutes a
basic weakness of the BTM. \ Even though the validity of golden-rule
transition rates in BTM is not generally assured, it has an enormous
acceptance in the literature. \ The toy free-electron MTJ theory, though
founded directly on a solution of the wave equation in the entire ideal
non-disordered FIF system having a flat barrier potential, was evaluated only
to leading order in the \textit{exponential }parameter $e^{-\kappa w}$
\cite{'89}. \ The BTM calculation for the same model agrees exactly with its
results, as one knew it should from previous non-spin dependent tunneling
theory \cite{Duke}.

Let us assume that BTM is correct to the same exponential degree for our
ideal-middle model as for the toy model. \ The previous section showed that
the BTM supports tunnel-polarization phenomenology in lowest order. Continuing
with BTM, we derive here a correction to the formulas (\ref{polfac}),
(\ref{OmegL}), and (\ref{OmegR}) for polarization which we find below varies
\textit{algebraically}, not exponentially, with $w^{-1}$. \ Therefore these
corrections should be reliable in spite of this general weakness of the BTM.

Further progress requires parametrization of the autocorellation functions
defined by Eq. (\ref{LM}). \ Note first the consequence of assuming that the
possibly disordered atomic configuration in F$_{\text{L}}$ produces no
electrostatic potential in F$_{\text{R}}$\ and \textit{vice versa}. \ From
Eqs. (\ref{lpsi}), (\ref{rphi}), and (\ref{LM}), in-plane translation of the
(disordered) microscopic potential of only the \textit{left electrode}
according to $\mathbf{s}\rightarrow\mathbf{s}+(B,C),$ where $(B,C)$ is a
periodic-lattice translation of the barrier middle,\ has the effects, from Eq.
(\ref{LM}), $\mathcal{L}_{\sigma}\rightarrow\mathcal{L}_{\sigma}%
\exp[i(\mathbf{k}^{\prime}\mathbf{-k})\cdot(B,C)]$ and $\mathcal{M}%
_{\sigma^{\prime}}\rightarrow\mathcal{M}_{\sigma^{\prime}}.$ Averaging over
all possible such phase changes makes $\mathcal{L}_{\sigma}$ and
$\mathcal{M}_{\sigma^{\prime}}$ diagonal and eliminates all terms with
$\mathbf{k\neq k}^{\prime}$ from the double sum in Eq. (\ref{Gam}). \ This
equation now becomes%
\begin{equation}
\Gamma_{\sigma,\sigma^{\prime}}=\frac{2\pi eV}{\hslash}\sum
\nolimits_{\text{\textbf{k}}}F^{2}(\mathbf{k})\mathcal{L}_{\sigma}%
(\mathbf{k})\mathcal{M}_{\sigma^{\prime}}(\mathbf{k})\label{Gamdiag}%
\end{equation}
using the now diagonal forms of $\mathcal{L}_{\sigma}$ and $\mathcal{M}%
_{\sigma^{\prime}}.$

Parenthetically, note that \textit{in the special case of vanishing disorder},
the state indices $p$ and $q$ become $m,\mathbf{k}$ \ and $n,\mathbf{k}$
respectively, with $m,n$ the respective band indices and \textbf{k }the
lateral crystalline momentum. Let the basis states be normalized to unity.
\ Then the diagonal elements of formulas (\ref{LM}) reduce to%
\begin{equation}
\mathcal{L}_{\sigma}=\Sigma_{m}|\lambda_{m,\sigma}(\mathbf{k})|^{2}%
/v_{x,m,\sigma}(\mathbf{k)}\text{ , \ }\mathcal{M}_{\sigma^{\prime}}%
=\Sigma_{n,\sigma^{\prime}}|\mu_{n,\sigma^{\prime}}(\mathbf{k})|^{2}%
/v_{x,n,\sigma}(\mathbf{k)}\label{nodisord}%
\end{equation}
with factors independent of $\sigma$ and $\sigma^{\prime}$ omitted. \ Here
$v_{x,m,\sigma}=\partial\varepsilon_{m,\sigma}(\mathbf{k})/\partial k_{x}$ and
$v_{x,n,\sigma^{\prime}}=\partial\varepsilon_{n,\sigma^{\prime}}%
(\mathbf{k})/\partial k_{x}$ are velocity components normal to the junction
plane. \ Their presence in these formulas follows from the restriction on
$\Sigma^{\prime}$ in the basic formula (\ref{tuncur}).

To evaluate Eq. (\ref{Gamdiag}) \textit{for finite disorder}, specialize to
small $V$ and constant $\mathcal{U}$ inside $\mathcal{B}.$ \ After evaluation
of the integral in Eq. (\ref{F(k)}), it reduces to the form%
\[
\Gamma_{\sigma,\sigma^{\prime}}=f_{1}\sum\nolimits_{\text{\textbf{k}}}%
\kappa^{2}(\mathbf{k})e^{-2w\kappa(\mathbf{k})}\mathcal{L}_{\sigma}%
(\mathbf{k})\mathcal{M}_{\sigma^{\prime}}(\mathbf{k})
\]
where $f_{1}$ does not depend on $\sigma$ or $\sigma^{\prime}.$ \ For large
$w,$ this sum weights small $k$ heavily, as mentioned above. \ Therefore
parametrize $\mathcal{L}_{\sigma}$ and $\mathcal{M}_{\sigma^{\prime}}$ for
small $k$ with the lateral spatial \textit{correlation scales} ($\xi_{\sigma}%
$,$\eta_{\sigma^{\prime}})$ defined by the formulas%
\begin{align}
\mathcal{L}_{\sigma}(\mathbf{k}) &  =\mathcal{L}_{\sigma}(0)[1-\xi_{\sigma
}^{2}k^{2}+\mathcal{O}(k^{4})],\text{ }\label{corscal}\\
\text{\ }\mathcal{M}_{\sigma^{\prime}}(\mathbf{k}) &  =\mathcal{M}%
_{\sigma^{\prime}}(0)[1-\eta_{\sigma^{\prime}}^{2}k^{2}+\mathcal{O}%
(k^{4})]\nonumber
\end{align}
and approximate Eq. (\ref{kap}) with $\kappa\approx\kappa_{0}+(k^{2}%
/2\kappa_{0})$ in the exponent of Eq. (\ref{F(k)}). \ After approximating
$\sum_{\text{k}}$ (over one BZ) with an infinite integral, one finds by
elementary integration a result equivalent, to first order in $w^{-1},$ to%
\begin{align}
\Gamma_{\sigma,\sigma^{\prime}}(w) &  \approx\Gamma_{\sigma,\sigma^{\prime}%
}^{(0)}(w)+\Gamma_{\sigma,\sigma^{\prime}}^{(1)}(w)\\
&  \approx f_{2}(w)\mathcal{L}_{\sigma}(0)\left(  1-\frac{\kappa_{0}%
\xi_{\sigma}^{2}}{w+\kappa_{0}^{-1}}\right)  \mathcal{M}_{\sigma^{\prime}%
}(0)\left(  1-\frac{\kappa_{0}\eta_{\sigma^{\prime}}^{2}}{w+\kappa_{0}^{-1}%
}\right)
\end{align}
where once again factors independent of both $\sigma$ and $\sigma^{\prime}$
are absorbed into $f_{2}.$ Thus to this approximation, $\Gamma_{\sigma
,\sigma^{\prime}}$ once again has the factored form (\ref{factoring}). \ (It
appears that in order $w^{-2},$ $\Gamma_{\sigma,\sigma^{\prime}}$ does not
separate this way into left- and right-dependent factors.) \ The corrected
left polarization factor, according to Eq.(\ref{polfac}) reduces upon
expansion to%
\begin{align}
P_{\text{L}} &  =P_{\text{L}}^{(0)}+\frac{1}{2}\left(  1-P_{\text{L}%
}^{(0)\text{ }2}\right)  \frac{\kappa_{0}(\xi_{-}^{2}-\xi_{+}^{2})}%
{w+\kappa_{0}^{-1}}+...\text{ \ \ with}\label{Pcorr}\\
P_{\text{L}}^{(0)} &  =\frac{\mathcal{L}_{+}(0)-\mathcal{L}_{-}(0)}%
{\mathcal{L}_{+}(0)+\mathcal{L}_{-}(0)},\nonumber
\end{align}
and similarly for $P_{\text{R}}.$ Thus, from given Bardeen basis functions one
can obtain polarization factors, correctly to order $w^{-1},$ in a disordered
electrode-barrier combination.

\section{Discussion}

Although it is valid only in the limit of weak transmission, predictions from
Bardeen's tunneling theory \cite{Bard} are interesting because it does not
require electron momentum within the electrodes to be conserved. \ Our
application to elastic tunneling through ordered or disordered magnetic
tunneling junctions yields these new conclusions:

\begin{itemize}
\item In Section 3 we found that the torque at constant external
\textit{voltage }is generally proportional to $\sin\theta$ [Eq.(\ref{TR=sin})]
\textit{. \ }This result is a direct consequence of the single-transition
nature of tunneling and the simple form of the spinor transformation
(\ref{tunmat}). \ It contrasts with the more general angular dependence
conditioned on electron structure and spin-channel resistance parameters in
the case of a metallic spacer \cite{'02}.

\item In general, polarization factors do not exist in the absence of special
assumptions, in agreement with previous theory \cite{MoodMatRev,TsymbRev}.

\item In Section 4 we found that if the polarization factors \textit{are} well
defined, then at constant applied voltage the electric current and in-plane
torque obey the relations (\ref{magcur}), (\ref{TR=sin}), and (\ref{TL=sin}).
\ These similar relations are inter-connected by the presence of the common
factor $J_{0}(V)$ which we do not attempt to calculate. \ The dimensionless
coefficients in these relations are expressed in terms of the polarizations by
$\tau_{\text{R}}=P_{\text{L}},$ $\tau_{\text{L}}=P_{\text{R}}$, $\iota
=P_{\text{L}}P_{\text{R}},$ implying $\iota=\tau_{\text{L}}\tau_{\text{R}}.$
\ In particular, these general relations are satisfied by the special results
of a direct solution of the Schroedinger equation for the toy model of
parabolic bands and ideal rectangular potential barrier \cite{'89}.

\item Experimentally, TMR is known to usually diminish with increasing
external voltage $V$ \cite{MoodMatRev,MiyazakiReview}. \ In Section 5 we
considered that it is the polarizing factor of the collector electrode which
decreases more strongly with $V$, resulting in the unsymmetric schematic
pattern of voltage dependence of torque indicated in Fig. 3. \ This lack of
symmetry due to the relations $\tau_{\text{R}}=P_{\text{L}}$ and\ $\tau
_{\text{L}}=P_{\text{R}}$ implies that the threshold voltage for initiation of
dynamic excitation will be increasingly asymmetric at the higher values (%
$>$%
100 mV) likely needed for writing in memory. \ Cases may well arise in which
voltage-driven switching works in only one direction. \ For selected
experimental junctions, switching is observed at a voltage sufficiently high
for TMR to become negligible \cite{Fuchs}. \ Our Fig. 3 indicates how this may
happen for switching in but one direction, from AP to P. \ However, our theory
would \textit{not} explain any \emph{symmetric} persistence of switching at
voltages great enough to destroy TMR, if this is observed.

\item Our approach to the validation of polarization factors complements
previous studies which accounted for atomic disorder in the barrier assuming
electrode states with well-defined crystalline momentum
\cite{TsymPet,MathUmer,Belash}. \ We assume that the barrier is thick and
includes an ideal crystalline or vacuum middle region of thickness $w$ as in
Fig. 4.  Then a newly derived polarization factor, given by Eq. (\ref{Pcorr}),
is valid to first order in $w^{-1}$ even in the presence of disorder in the
electrodes and interfaces sufficient to destroy the conservation of lateral
crystalline momentum throughout the electrode and interface regions. \ The key
basis-state weight factors (\ref{OmegL}) and (\ref{OmegR}) are more general
than the conventional local state density.

\item Our conclusion that the validity of polarization factors increases with
increasing $w$ tends to undermine our predictions of voltage asymmetry of
torque shown schematically in Fig. 3. \ For, experimental spin-transfer
effects such as switching will require very thin barriers, making the
separability condition assumed in Fig. 3 less valid. \ Previous proposals
\cite{TsymbRev,Belash} that validity of polarization factors is attributable
to unique defect states or amorphicity in the barrier are more promising in
this respect.

\item Belashenko and co-authors \cite{Belash} find that certain
first-principle TMR computations for realistic barrier thickness may be poorly
approximated by proportionality to $e^{-\kappa w}$. \ This casts additional
doubt on the applicability of the ideal middle to the very thin junctions
needed for spin-momentum transfer experiments. \ However, our conclusions from
this model may bear significantly on magneto-resistance experiments carried
out with greater thickness, as suggested below.

\item Our parametrized expression (\ref{Pcorr}) for dependence of tunnel
polarization on ideal-middle thickness \emph{w} is without precedent. \ A
strong dependence is expected from certain compositions, like Co, Ni, and
certain alloys, such as FeCo, lying on the negative-slope region of the
Slater-Pauling curve \cite{SPCurv}; for, their strong contrast between heavily
4sp-weighted density of majority-spin and heavily 3d-weighted density of
minority-spin bands may be reflected in strongly contrasting magnitudes of
left lateral autocorellation scales $\xi_{+}$ and $\xi_{-}.$ \ Theoretical
estimation of the left polarization factor will require prior first-principle
computation of the Bardeen basis functions $\psi_{p,\sigma}$ for the
disordered electrode-barrier system. \ From these, one must invert the series
(\ref{lpsi}) to evaluate the diagonal elements of the Fourier coefficients
$\lambda_{p,\sigma}.$ \ Then Taylor expansion of the diagonal element in the
first Eq. (\ref{LM}) for substitution into the first Eq. (\ref{corscal})
provides the coefficients $\mathcal{L}_{\pm}(0)$ and $\xi_{\pm}.$ \ These
parameters must then be substituted into Eqs. (\ref{Pcorr}) to obtain the left
polarization factor.

\item In fact, experimental junctions having composition Fe/Al$_{2}$O$_{3}%
$/FeCo show dependence of TMR on barrier thickness\textit{ }%
\cite{MiyazakiReview} at T=2 K where our assumption of elastic tunneling
should be valid. \ A monotonic dependence on thickness, expected from Eq.
(42), is observed for two crystallographic orientations on single-crystal Fe,
but not on the third. \ Although the (say) right electrode (FeCo) lies on the
negative-slope side, the left electrode (Fe) lies on the positive-slope side
of the Slater-Pauling curve where high 3d density exists for both\ signs of
spin so that there may be little difference between $\xi_{+}$ and $\xi_{-}$.
\ Junctions with both electrodes taken from the negative-slope side may yield
a more pronounced thickness dependence of TMR on barrier thickness according
to the present theory.
\end{itemize}

\textbf{Acknowledgments. \ }The author is grateful to G. Mathon for a related
preprint, and for helpful discussions with W. Butler, E. Tsymbal, K.
Belashchenko, M. Stiles, Y. Bazaliy, J. Sun, S. S. P. Parkin, P. Nguyen, G.
Fuchs, D. Worledge, and P. Visscher.

\end{document}